\documentstyle[preprint,epsf,aps,amsmath,amsfonts,amssymb]{revtex}
\draft
\tighten
\begin{document}
\begin{sloppypar}

\title{Algebraic approach in the study of time-dependent nonlinear integrable 
systems: Case of the singular oscillator}

\author {Jayendra N. Bandyopadhyay\footnote{jayendra@prl.ernet.in},
A. Lakshminarayan and
Vijay B. Sheorey}
\address{Physical Research Laboratory,\\
Navrangpura, Ahmedabad 380 009, India.}

\date{\today}

\maketitle

\begin{abstract}
  
  The classical and the quantal problem of a particle interacting in
  one-dimension with an external time-dependent quadratic potential
  and a constant inverse square potential is studied from the
  Lie-algebraic point of view. The integrability of this system is
  established by evaluating the exact invariant closely related to the
  Lewis and Riesenfeld invariant for the time-dependent harmonic
  oscillator. We study extensively the special and interesting case of
  a kicked quadratic potential from which we derive a new integrable,
  nonlinear, area preserving, two-dimensional map which may, for
  instance, be used in numerical algorithms that integrate the
  Calogero-Sutherland-Moser Hamiltonian.  The dynamics, both classical and
  quantal, is studied via the time-evolution operator which we
  evaluate using a recent method of integrating the quantum
  Liouville-Bloch equations \cite{rau}.  The results show the exact
  one-to-one correspondence between the classical and the quantal
  dynamics. Our analysis also sheds light on the connection between
  properties of the $SU(1,1)$ algebra and that of simple dynamical
  systems.
  
\end{abstract}


\newpage

\section{Introduction}

The classical and quantum mechanical study of time-dependent
Hamiltonian systems are generic and important. They span a wide
spectrum of subjects ranging from interaction between atoms and
radiation \cite{sonza,lo}, adiabatic \cite{berry1,shap} and
non-adiabatic \cite{wang1,moore} Berry phase, time-dependent harmonic
oscillators \cite{lewis1,lewis,lr} and quantum motion of a particle
in a Paul trap \cite{paul,brown}, to time-dependent mean-field theory
(cranking model \cite{beng}) in particular and the time-dependent shell 
model \cite{ayik} in general.

The question of the existence of invariants (constants of motion) is
one of  central importance in the study of any dynamical system, be
it classical or quantal. A basic theorem of classical mechanics 
asserts that if the number of independent invariants, satisfying 
certain conditions, is equal to the number of degrees of freedom, 
then the motion can be reduced to quadratures, or equivalently, 
an action-angle transformation to a Hamiltonian dependent only on 
the actions can be found \cite{Arnold}. Such systems are integrable.

The lack of sufficient number of invariants invariably leads to a
phase space that has a nonzero measure of {\it chaotic} trajectories
\cite{Licht}. For time-independent Hamiltonian systems the Hamiltonian
itself is an invariant. However, when the Hamiltonian is an explicit
function of time, it is no more an invariant, and this is of course a
reflection of the non-conservation of energy. Various methods have
been used to obtain approximate solutions for time-dependent problems,
e.g., the adiabatic approximation, the sudden approximation,
time-dependent perturbation techniques, etc.

The most widely studied time-dependent Hamiltonian system is the
time-dependent harmonic oscillator (TDHO). It has long been a problem of
considerable interest because of its varied applications in different
areas of physics, for instance in molecular physics, quantum
chemistry, quantum optics and  plasma physics. 
The Hamiltonian of this system is given by,
\begin{equation}
{\cal H}(t) \,=\, \frac{1}{2}\, p^2 \,+\, \frac{1}{2}\, \omega^2(t) \,q^2,
\label{ham1}
\end{equation}
\noindent where $p$ and $q$ are conjugate canonical variables.  
The adiabatic invariant for this system was originally given at the
first Solvay Congress in 1911 when the Hamiltonian of this system was
used as an approximate Hamiltonian for the slowly lengthening
pendulum \cite{solvay}.  

The study of this one-dimensional TDHO was greatly
advanced due to the work of Lewis \cite{lewis1,lewis} and Riesenfeld
\cite{lr}. Lewis
\cite{lewis1} determined the exact invariant by applying Kruskal's asymptotic
method \cite{krus} and showed that a previously known adiabatic invariant was
in fact an exact invariant. 
Later Lewis and Riesenfeld (LR) \cite{lr} determined
that same invariant by starting with the assumption of the existence of an 
explicitly time-dependent, homogeneous and quadratic invariant of the form given
by, 
\begin{equation}
I(t) \,=\, \frac{1}{2} [\alpha(t) \,p^2 + \beta(t) \,q^2 + 2 \,\gamma(t) \,p\,q ],
\label{inv1}
\end{equation}
\noindent where the coefficients $\alpha(t), \beta(t)$ and $\gamma(t)$ are 
time-dependent real functions  and $I(t)$ satisfies the condition,
\begin{equation}
\frac{dI}{dt} \,\equiv\, \frac{\partial I}{\partial t} \,+\, \{I,{\cal
  H}(t)\}
\, =\, 0.
\label{cond}
\end{equation}
\noindent Here $\{ *, * \}$ denotes the usual Poisson bracket. From the above
two equations and after some calculations they derived the exact
invariant as,
\begin{equation}
I(t) \,=\, \frac{1}{2}\, \left[ \rho^2 \,p^2 \,+\, \left( \dot{\rho}^2 \,+\, 
\frac{1}{\rho^2} \right) \,q^2 \,-\, 2 \,\rho \,\dot{\rho} \,p\,q \right], 
\label{inv2}
\end{equation}
\noindent with $\rho(t)$ satisfying the subsidiary condition
\begin{equation}
\ddot{\rho}\, +\, \omega^2(t) \,\rho - \frac{1}{\rho^3} \,=\, 0. \label{coef}
\end{equation}
\noindent This more complicated, nonlinear, differential equation
represented an advance due to the reason that {\it any} particular
solution $\rho(t)$ of the above equation would give the exact
invariant $I(t)$ for {\it all} initial conditions of $p$ and $q$. 

The above technique actually predates the work of Lewis and Riesenfeld in a 
slightly different context, and we briefly point this out.
We
note that the equation of motion of the TDHO is, $d^2 q/dt^2 \,+\,
\omega^2 (t) ~q \,=\, 0$.  This is of the same form as the
time-independent, one-dimensional Schr{\"o}dinger equation, if we
assume that $t$ represents the spatial coordinate and $q$, the
wave-function. According to an early work by Milne (1930) \cite{milne},
it is possible to solve this Schr{\"o}dinger equation using the
knowledge of any particular solution of the Eq.~(\ref{coef}), after proper
identifications. 
Therefore, any particular solution of the subsidiary condition
can also give the exact solution for the TDHO.

Lewis \cite{lewis} attempted to give an interpretation of $I(t)$ as
the most general homogeneous quadratic invariant possible for the
Hamiltonian of the one-dimensional TDHO. A more natural and physical
interpretation has been suggested by Eliezer and Gray \cite{eli} in
terms of two-dimensional auxiliary motion, i.e., in terms of a
two-dimensional uncoupled TDHO. They showed that the above subsidiary
condition Eq. (\ref{coef}) is the radial equation of motion for this
two-dimensional system and the invariant $I(t)$ is proportional to the
conserved angular momentum of this auxiliary motion. G\"{u}nther and
Leach \cite{gun} interpreted $I(t)$ in terms of canonical
transformations and under their transformation the invariant $I(t)$
became the Hamiltonian of the one-dimensional time-independent
harmonic oscillator of unit frequency.
 
Besides these previous interpretations of $I(t)$, we can interpret the
form of $I(t)$ chosen by Lewis and Riesenfeld (LR) \cite{lr} from the
Lie-algebraic point of view. The Hamiltonian of the one-dimensional
TDHO is formed by the dynamical variables $\frac{1}{2} p^2$ and
$\frac{1}{2} q^2$. These two dynamical variables together with $p\,q$
are generators of the closed $SU(1,1)$ algebra under the Poisson
bracket operation and $I(t)$ was chosen in \cite{lr} as the linear
combination of these generators. Generalizing the LR-invariant from the
Lie-algebraic point of view, we expect that if any time-dependent
Hamiltonian be the combination of the generators of any closed
algebra, an invariant would be a linear combination of the generators
of that closed algebra with time-dependent coefficients.

The integrability of the one-dimensional TDHO is not surprising,
because this is a linear system. We know, one-dimensional
time-dependent Hamiltonians usually lead to non-integrability, e.g., a
simple pendulum whose length varies in time. Except for the adiabatic or
small oscillation approximations, this was the problem posed by
Lorentz at the above mentioned Solvay congress and the solutions have
the possibility of displaying chaos.

Now a natural question is if there exist one-dimensional
time-dependent nonlinear Hamiltonians which are also integrable? In
fact, the singular oscillator with a centrifugal force potential
provides an important example whose kinematics is still within the
$SU(1,1)$ algebra. In the second section, we have discussed such a
nonlinear Hamiltonian and we have also derived its invariant from our
new Lie-algebraic interpretation of the LR-invariant. Using this
Hamiltonian we have constructed an integrable, area preserving,
nonlinear map and derived its exact invariant with a knowledge of
its fixed points. 

We have determined the classical time-evolution (Perron-Frobenius)
operator following a recent method elucidated by Rau \cite{rau}. This
method was originally used for solving the quantum Liouville-Bloch
equation, but we have applied this in the classical case and have been
able to show the one-to-one correspondence between this nonlinear map
and the linear map (derived from the Hamiltonian of the
one-dimensional TDHO). In the third section, we have studied the
quantum dynamics of the nonlinear map and again we have used Rau's
method to determine the quantum time-evolution operator. This
identical approach to classical and quantum dynamics helps us to show,
rather trivially, the exact one-to-one correspondence between the
classical and the quantal.  Our study also has the dimension of
interpreting the mathematics of the $SU(1,1)$ algebra from a dynamical
systems viewpoint. For instance, we derive powers of non-commuting
products of exponentials of the $SU(1,1)$ algebra. This reveals when
the powers can degenerate to the identity, something that is
immediately clear from the dynamics of the underlying system due to
the presence of degenerate periodic orbits.

\section{ The classical kicked  singular oscillator} 

First, we will describe the general Hamiltonian of which an
important special case constitutes the rest of the paper.
For the existence of an invariant for the one-dimensional TDHO, the
subsidiary condition Eq. (\ref{coef}) has to be integrated with {\it some}
initial conditions, that is we should be able to determine a
particular solution. The subsidiary condition, given in Eq. (\ref{coef}),
is a nonlinear, time-dependent, equation and its integrability is not
immediately obvious. If we assume $\rho$ as a position variable, say
$q$, then Eq. (\ref{coef}) is the equation of motion corresponding to the
Hamiltonian given by, 
\begin{equation}
{\cal H}(t)\, = \,\frac{1}{2}\,\left( p^2 \,+\, \frac{k}{q^2} \right)
\,+\,
 \frac{1}{2} \,\omega^2(t) \,q^2,
\label{ham2}
\end{equation}
with $k=1$.  This is a time-dependent Hamiltonian,
and has also been studied for long: it 
was studied in part by Lewis and Leach \cite{lewis2}, Camiz {\it et. al.}
and Pedrosa {\it et. al.} \cite{camiz} studied this Hamiltonian 
quantum mechanically. 
The new nonlinear force in the system is a {\it `Centrifugal force'} and it
appears in many integrable systems, including the celebrated
Calogero-Sutherland-Moser \cite{calo,suth,moser} many-body
Hamiltonian. 

The Hamiltonian in Eq. (\ref{ham2}), is formed by the dynamical
variables $\frac{1}{2}\left( p^2 \,+\, \frac{k}{q^2} \right)$ and
$\frac{1}{2}\,q^2$. These two variables together with $p\,q$, also
form the $SU(1,1)$ closed algebra.  Following our algebraic
interpretation of $I(t)$, we can assume the invariant of this
nonlinear Hamiltonian to be of the form,
\begin{equation}
I(t) \,=\, \frac{1}{2}\,\alpha(t)\left( p^2 \,+\, \frac{k}{q^2}
\right) \,+\,
 \frac{1}{2}\,\beta(t) \,q^2 \,+\, 2 \, \gamma(t) \,p \,q. \label{inv3}
\end{equation}
\noindent Again, by the same substitutions and identical procedures we 
get
\begin{equation}
I(t) \,=\, \frac{1}{2}\,\rho^2 \left( p^2 \,+\, \frac{k}{q^2} \right)
\,+\,
 \frac{1}{2} \,\left( \dot{\rho}^2 \,+\, \frac{1}{\rho^2} \right) \,q^2 \,-\, \,\rho\,\dot{\rho} \,p \,q, 
\label{inv4}
\end{equation}
\noindent with the same subsidiary condition as given in
Eq. (\ref{coef}). It is also possible to determine this invariant
using time-dependent canonical transformations \cite{pedrosa},
however, our algebraic method accomplishes this more elegantly.  In
this case, the equation of motion and the subsidiary condition are the
{\em same} nonlinear equation, but the fact that we need only one
particular solution of this equation to determine the invariant shows
the power of the methodology and the integrability of the above
Hamiltonian in Eq.  (\ref{ham2}). In Fig. \ref{strobo} we show a
special case corresponding to $\omega^2(t)\,=\, 1 + \cos(\sqrt{2}\,t)$
where the existence of the invariant is reflected in the regular
structures.

\subsection{The integrable discrete system}

Now we use the above to construct a nonlinear integrable map.
Integrable discrete systems, or maps, have been intensely studied for
some time now and many interesting methods and results have been
found. Integrable maps are important for the following reasons.
Firstly, studying the dynamics of a mapping is more simpler than the
dynamics of a continuous system since it involves direct iteration.
Secondly, if one wants to study numerically any integrable continuous
system, it is absurd to use non-integrable numerical schemes which
destroy the basic properties of the system. Therefore, for numerical
studies of any integrable system, one has to find a discrete
integrable version of the system. The discrete map we discuss below, or
extensions thereof, for instance, may be used in the numerical studies
of the Calogero-Sutherland-Moser model. Lastly, we can argue that the discrete
systems are more fundamental ones since they contain continuous ones
as special limits.

Studying the stroboscopic map of any time-periodic system is always
possible, at least numerically. To get an analytical expression a much
studied form of the time dependence is: $\omega^2(t) = \omega^2 \,T\,
\sum_{n} \delta (t - nT)$. The standard map and its quantization as
well as an experimental realization of the time-dependence through  a
periodically pulsed laser field \cite{raizen} provides a well 
known example of this kind. The map corresponding
to the TDHO is a linear map, which is very simple classically. 

Here we will study the map corresponding to the  nonlinear system. 
The Hamiltonian of interest is, 
\begin{equation}
{\cal H}(t) \,=\, \frac{1}{2} \,\left( p^2 \,+\, \frac{k}{q^2} \right)
\,+\, \frac{1}{2}\,
\omega^2 \,T \,q^2 \sum_{n} \delta (t - nT), \label{ham6}
\end{equation}
\noindent and the corresponding Hamilton's equations of motion are,
\begin{eqnarray}
\dot{p} &=& \frac{k}{q^3} - \omega^2 \,T \,q \sum_{n} \delta (t - nT), 
\nonumber \\
\dot{q} &=& p.
\end{eqnarray}
\noindent The equation of motion of the system is given by,
\begin{equation}
\ddot{q} \,+\, \omega^2 \,T \,q \sum_{n} \delta (t - nT) \,-\,
\frac{k}{q^3} \,=\, 0.
\label{same}
\end{equation}
\noindent This equation is the same as that of the subsidiary
condition given in Eq. (\ref{coef}), except for the constant $k$.
Integrating Hamilton's equations of motion from just after the $n$-th kick
to just after the $(n+1)$-th kick and defining new scaled variables
\[ p \rightarrow
k^{\frac{1}{4}} \,T^{- \frac{1}{2}} \,p,\;\;\;  \mbox{and}\;\;\;
q \rightarrow k^{\frac{1}{4}} \,T^{\frac{1}{2}} \,q,\]
the phase space map of the system is, 
\begin{eqnarray}
q_{n+1} &=& \sqrt{p_{n}^2 + \frac{1}{q_{n}^2} + q_{n}^2 + 2 p_{n} q_{n}},
\nonumber\\
p_{n+1} &=& \frac{p_{n}^2 + \frac{1}{q_{n}^2} + p_{n} q_{n}}{q_{n+1}}
- \Omega^2  q_{n+1}, \label{map}
\end{eqnarray}
\noindent where $\Omega = \omega \,T$.

Note that the scaling has removed the $k$ dependence, but that the
scaling is singular at $k=0$ and in fact this limit leads to the
removal of the $1/q^2$ term, as is discussed further below.  If we
plot this phase space map numerically, it clearly shows its
regular behaviour, as illustrated in Fig. \ref{pq}. This rather
complicated looking discrete map has a very simple behaviour,
reflecting the fact that both this and a linear map, derivable from a
TDHO, have common algebraic antecedents. This linear map is derived
from the Hamiltonian in Eq. (\ref{ham6}) by simply discarding the
$1/q^2$ term (the $k=0$ limit)  and is given by
\begin{eqnarray}
\label{limap}
q_{n+1} &=& p_n + q_n, \nonumber \\ 
p_{n+1} &=& p_n - \Omega^2  \,q_{n+1}.  
\end{eqnarray} 
\noindent In fact
the property of the linear map that exclusively quasi-periodic or
entirely periodic behaviour exists for different values of the
parameter is also observed for the nonlinear map. For $0< \Omega^2<4$
the motion is bounded and stable and for all other values the motion
is unbounded and unstable, in both the maps.  The invariant in the
case of the linear map describes either an ellipse or a hyperbola, now
we determine the invariant of the nonlinear map using the method of
LR.

Using Eq. (\ref{inv4})  the invariant of this map would be of the form,
\begin{equation}
I \,= \,\rho^2 \left( {p_n}^2 \,+\, \frac{1}{{q_n}^2} \right) \,+\,
 \left( \dot{\rho}^2 \,+\, 
\frac{1}{\rho^2} \right) {q_n}^2 \,-\, 2 \,\rho\,\dot{\rho} \,{p_n} \,{q_n}, 
\label{inv5}
\end{equation}
\noindent where $\rho$ satisfies the same subsidiary condition as given in 
Eq. (\ref{coef}), but now $\omega^2(t) = \omega^2 \,T \,\sum_{n} \delta ( t
- nT )$. If we are able to
determine any particular solution of the subsidiary condition, then we
can use that solution to get the invariant of the map Eq.  (\ref{map}). 
As we have already mentioned, the equation of motion of
this nonlinear kicked system is the same as that of the subsidiary
condition for $\rho$, as shown in Eq. (\ref{same}), therefore if we are able
to get {\it any} solution of the nonlinear map Eq. (\ref{map}),
that solution should also be the solution for $\rho$. The most simple
solution corresponds to the fixed-point. Therefore the problem of
determining the invariant has been reduced to determining the fixed-point of 
the nonlinear mapping. The fixed point is at,
\begin{eqnarray}
q^{\ast} &=& - \frac{2}{\Omega^{1/2} 
\left[ 4 \left( 4 \,-\, \Omega^2 \right) \right]^{1/4}}, \nonumber \\ 
p^{\ast} &=&
\frac{\Omega^{3/2}}{\left[ 4 \left( 4 \,-\, \Omega^2 \right) \right]^ {1/4}}. 
\end{eqnarray}
This is a particular solution of $\rho$ and $\dot{\rho}$. Note that
unlike the linear map whose fixed point at the origin is independent
of the system parameters, the fixed points here move with the
parameter and as the system approaches instability ($\Omega^2
\rightarrow 4$), they  approach infinity. Using these solutions, we get the
invariant of the mapping within an arbitrary multiplicative constant as,
\begin{equation}
I \,=\, p_{n}^2 \,+\, \frac{1}{q_{n}^2} \,+\, \Omega^2 \,\left(
  q_{n}^2 \,
+\, q_n \,p_n \right).
\label{inv6}
\end{equation}
The lack of a time subscript on $I$ indicates its constancy. This
invariant is valid even when the motion is classically unstable.

\subsection{ The classical evolution operator}

We would like to study the nonlinear map from a point of view that
simultaneously explains its dynamics as well as sets the stage for
quantum mechanical work. Here we introduce the classical time-evolution
operator for studying the dynamics of the system. For a given
dynamical variable, say $V$, the corresponding Liouville operator is
denoted by $L_V$ and it is defined as $L_V \equiv \{V, *\}$, where
$\{*, *\}$ denotes the usual Poisson bracket.

We know the dynamical equation for any dynamical variable, say $f$, is given by,
\begin{equation}
\frac{df}{dt} \,=\, - \{H, f\} \,\equiv \,-\, L_H f \label{dfdt},
\end{equation}
\noindent where $H$ is the Hamiltonian of the system and $L_H$ is the Liouville
operator corresponding to $H$. Let us define $A \equiv \frac{1}{2} (p^2 + \frac
{k}{q^2}), B \equiv \frac{1}{2} \,q^2$ and $\{A, B\} = - p\,q \equiv
- 2\,C,$ i.e., $C = - \frac{1}{2} \,p\,q.$ The triad $(A, B, C)$ form the 
closed $SU(1, 1)$ algebra. Then we can write,
\begin{equation}
L_H  = \left[ L_A + \omega^2 \,T \,L_B \sum_{n} \delta (t - nT) \right], 
\end{equation}   
\noindent where $L_A$ and $L_B$ are the Liouville operators corresponding to
$A$ and $B$ respectively. These Liouville operators together with $L_C$, the
Liouville operator corresponding to $C$, forms the same $SU(1,1)$ algebra as
that of $(A, B, C)$ but under the Lie bracket operation, i.e., $[L_A, L_B] =
- 2 \,L_C, [L_C, L_A] = L_A$ and $ [L_C, L_B] = - L_B$.

From Eq. (\ref{dfdt}), the dynamical equation for the classical time-evolution 
of any arbitrary function $f$ on the  phase space  would be,
\begin{equation}
\frac{df}{dt} \,=\, - L_H \,f\,=\, - \left[ L_A \,+\, 
\omega^2 \,T \,L_B \sum_{n} \delta (t - nT) \right]\,f. 
\label{dFdt}
\end{equation}
\noindent Integrating the above equation Eq. (\ref{dFdt}) in between the time 
$t = 0$ to $t = T$, we get the classical
time-evolution operator from just after zero time to just after the 
first kick.
We write $f\left(q\,(t=T),p(t=T)\right)=F \, f\left(q\,(t=0),\,p\,(t=0)
\right)$ where
\begin{equation}
F \,=\, \exp\left( -\omega^2 \,T \,L_{B} \right) \exp\left( -L_{A} \,T \right). 
\label{flou}
\end{equation}
\noindent This Perron-Frobenius operator can be regarded as the 
classical Flouquet operator.  To understand the dynamics for time $n$,
we have to determine the power $F^{n}$. The one parameter Abelian
group of the powers completely specifies the dynamics at all time.
However, $F$ is itself a product of the exponential of two
non-commuting operators which do not even commute with their
commutator $L_C$. We can write $F$ in a single exponential form,
following an earlier work of Truax \cite{truax}, but that form is so involved
that it becomes difficult to extract the essentially simple dynamics
of the system.

We apply a recent operator method, referred to earlier \cite{rau}, to
derive the classical time-evolution operator at any time.  The general
procedure of this method is simple to describe. If we have any general
time-dependent Hamiltonian of the form given by,
\begin{equation}
{\cal H}(t) \,=\, \sum_{i=1}^{n} a_{i}(t) H_{i},
\end{equation}
\noindent where $(a_{i}(t), i=1,\ldots,n)$ are a set of linearly independent 
general complex function of time and the dynamical variables $(H_{i},
i=1,\ldots,n)$ are the generators of any $n-$dimensional closed Lie
algebra. Corresponding Liouville operators $(L_{H_i}, i=1,\ldots,n)$
would also form that same algebra.  Then the classical time-evolution
operator $F(t)$ can be expressed in the product form :
\begin{equation}
F(t) \,=\, \prod_{j=1}^{n} \exp [ b_{j} (t) \,L_{H_j}].
\end{equation}
\noindent Therefore, in the classical case we can start with 
the time-evolution 
operator of the form : 
\begin{equation}
F(t) \,=\, \exp[X(t)\,L_{B}]\;\exp[Y(t)\,L_{C}]\;\exp[Z(t)\,L_{A}], 
\label{Ft}
\end{equation}
\noindent where $X(t), Y(t)$ and $Z(t)$ are real functions of time. 
From the initial condition $F(0)={\bf 1}$, we
have $X(0) = Y(0) = Z(0) = 0$. Now, substituting this product form of
$F(t)$ in Eq. (\ref{dFdt}), and repeatedly applying the 
Campbell-Baker-Hausdorff
(CBH) formula we can cast it into a form such that $F(t)$ is
pushed to the extreme right in the LHS of the equation 
Eq. (\ref{dFdt}). This
yields a set of first order differential equations for the introduced
functions of time :
\begin{eqnarray}
\label{clco}  
\dot{X} &=& - X^2 - \omega^2 \,T \,\sum_{n} \delta (t - nT), \nonumber \\    
\dot{Y} &=& 2 \,X, \\
\dot{Z} &=& - e^{-Y}. \nonumber
\end{eqnarray}
\noindent In the equation Eq. (\ref{Ft}) we can choose the exponential 
operators in different orders, but we found that this leads to  sets of
differential equations whose solutions may not even exist for such
kicked systems. 

We now treat the case of delta kicks. Integrating the above
equations in between two consecutive kicks and defining 
\[ x = T \, X,\; \;y = Y, \;\; \mbox{and} \; \; z =\frac{Z}{T}, \] 
we  get a nonlinear mapping, a ``coefficient'' mapping for the new 
dimensionless variables $(x, y, z)$:
\begin{mathletters}
\label{xyz}
\begin{eqnarray}
x_{n+1} &=& \frac{x_{n}}{1 + x_{n}} - \Omega^2, \label{xyza}\\  
y_{n+1} &=& y_{n} - \log \left[ (1 + x_{n})^2 \right], \label{xyzb}\\
z_{n+1} &=& z_{n} + \frac{1}{\Omega^2} ( x_{n+1} - x_{n}).\label{xyzc}
\end{eqnarray}
\end{mathletters}
To be explicit the $n$-th power of the operator $F$ is
\begin{equation}
\label{powerF}
F^n \,=\,  \exp[\,x_n \,T \,L_{B}]\;\exp[\,y_n \,L_{C}]\;\exp[\,z_n \,L_{A}/T]. 
\end{equation}
\noindent From the initial condition $F^0 \,=\, {\bf 1}$, we have $x_{0} \,=\, 
y_{0} \,=\, z_{0} \,=\, 0$. The time development is now entirely buried in 
the scalar functions $x_n,y_n,$ and $z_n$.

The most important of the recursion equations in Eq. (\ref{xyz}) is
the first one.  We note that this equation viewed as a transformation
is a special case of the ``bilinear'' conformal transformation in
complex analysis or a special case of the projective group $PSL(2,{\Bbb R})$.
We solved this nonlinear map by constructing an auxiliary {\it two
dimensional } linear map.  This is not entirely surprising as
lurking behind the one dimensional nonlinear singular oscillator is a
two dimensional linear one.  This gives a new insight into the often
stated close relationship between the harmonic oscillator and the
singular oscillator.
 
Before we solve these equations explicitly we point out the following
interesting fact.  Consider the Hamiltonian
\begin{equation}
H(q,p,t)\,=\, q \,p^2 \, +\, \Omega^2 \,q \,  \sum_{n} \delta (t - n),
\end{equation}
written in terms of dimensionless canonical variables.  The triad $(q
\, p^2, \, q,\, q \, p )$ form the algebra $SO(2,1)$ under the
Poisson bracket operation and this algebra is locally isomorphic to
$SU(1,1)$ \cite{perelo,cornwell}. Then the resulting map for $q_n$ and $p_n$ is 
\begin{mathletters}
\begin{eqnarray}
p_{n+1} &=& \frac{p_{n}}{1 + p_{n}} - \Omega^2,\\  
q_{n+1} &=& q_n  (1 + p_{n})^2. 
\end{eqnarray}
\end{mathletters}
Thus if we identify $x_n \equiv p_n$ and $y_n \, \equiv \, -\log(q_n)$ we get 
the first two of the recursion relations in Eq. (\ref{xyz}). We note that the
$SO(2,1)$ algebra underlies the Morse oscillator \cite{palma}. 
Therefore the  coefficient map also has an algebraic 
basis and in fact one that is practically identical to the original
one.
  
We return now to solve Eq. (\ref{xyz}).  First we define $s_{n} \equiv
1 + x_{n}$, and in terms of $s_{n}$, Eq. (\ref{xyza}) becomes
\begin{equation}
s_{n+1} \,= \,\eta \,-\, \frac{1}{s_{n}}, \label{eta}
\end{equation}
\noindent where $\eta \,\equiv\, 2 \,-\, \Omega^2$, and 
from the initial condition $x_{0}=0$, we have $s_{0}=1$. 
Construct the auxiliary linear map:
\begin{equation}
\left( \begin{array}{c} a_{n+1}\\ b_{n+1} \end{array} 
\right) = \left( \begin{array}{cc} \eta & -1\\ 1 & 0 \end{array} \right) \left(
\begin{array}{c} a_n \\b_n \end{array} \right),
\end{equation}
with initial conditions $a_0\,=\,b_0\,=\,1$. 
We identify $s_n \equiv a_n/b_n$.

To get the general form of $s_n$, we have to diagonalize the matrix $M 
\equiv ((\eta, -1),(1,0))$. The 
eigenvalues of $M$ are,
\begin{equation}
\lambda_{\pm} = \frac{1}{2} \left( \eta \pm \sqrt{\eta^2 - 4} \right).
\end{equation}
\noindent Whether $\lambda_{\pm}$ is real or complex is dependent on
$\eta$. Therefore just as for the linear map
we have to study separately three different regions for the parameter $\eta$, 
these are : 

\vspace{0.3cm}

\noindent {\bf Case 1: $\eta^2 < 4$, i.e., $0 < \Omega^2 < 4$.}

\vspace{0.3cm}

In this case the eigenvalues of $M$ are complex-conjugate of each
other, {\it i.e.},  we can write 
\[ \lambda_{\pm} = \exp(\pm i \sigma), \; \; \mbox{ where}\; \; 
\sigma \equiv \cos^{-1}(\eta/2)\, =\,\cos^{-1}\left(\frac{2 -
\Omega^2}{2}\right).\]
  The dynamics is  stable and bounded.  After
diagonalizing the matrix $M$, we  determine $a_n$ and $b_n$ to  
finally get:
\begin{equation}
x_n \,=\, \cos\sigma \,-\, \sin\sigma \,\tan\left[ (2n-1) \frac{\sigma}{2}
\right] - 1.
\end{equation}
\noindent In terms of $x$ we can obtain the solution for $y$ and $z$ as :
\begin{eqnarray}
\label{yz}
y_n &=& - 2 \sum_{k = 0}^{n-1} \log \left| 1 + x_k \right|, \\
z_n &=&  \frac{x_n}{\Omega^2}. 
\end{eqnarray}

When $\sigma = 2\pi m/N$, where $m$ and $N$ are coprime integers, one 
sees, after some algebra, that
\[ x_n \,=\, x_{n+N}, \;\; y_n \,=\, y_{n+N}, \;\; z_n \,=\, z_{n+N}. \]
Therefore for these particular values of $\sigma$ (or of the
corresponding value of $\Omega$), the above three-dimensional map for
the coefficients is {\it exactly} periodic. 
To be explicit at these values of the parameter
\begin{equation}
\label{ident}
F^N={\bf 1},
\end{equation}
the Perron-Frobenius operator becomes unity at time $N$. The quantum 
equivalence to be discussed below will be complete and the 
Flouquet unitary operator will become unity at the same time.
Fig. \ref{xyz1} shows the coefficients for a periodic case.
Since the coefficient $z$ is proportional to $x$,  it would 
follow the same behaviour as $x$. 
This map in Eq. (\ref{map}) would also be periodic with the
same period, and corresponds to the phase space shown in Fig. \ref{pq}(a).
For other values of $\Omega$, this map is quasi-periodic. Equivalently,
the phase space map Eq. (\ref{map}) also displays the above behaviour
for corresponding values of $\Omega$. This shows that our operator
approach for studying the classical map is in one-to-one
correspondence with the phase space dynamics.
Fig. \ref{xyz2} shows the coefficients for a quasi-periodic case, 
corresponding to the phase space in  Fig. \ref{pq}(b). Again the behaviour
of $z$ is same as $x$.
\vspace{0.3cm}

\noindent {\bf Case 2: $\eta^2 > 4$, i.e., $\Omega^2 > 4$ or $\Omega^2 < 0$}.

\vspace{0.3cm}

\noindent We can divide this case into two parts. They are :

\vspace{0.3cm}

\noindent (a) $\eta > 2$, i.e., $\Omega^2 < 0$ :

\vspace{0.3cm}

\noindent In this part the eigenvalues of $M$ are real and positive. 
Therefore we can take 
\[ \lambda_{\pm} \,=\, \exp(\pm
  \sigma),\;\; \mbox{where}\;\; \sigma \equiv \cosh^{-1} (\eta/2) \,=\,
\cosh^{-1}\left( \frac{2 - \Omega^2}{2} \right). \] 
In this part the
dynamics is unstable and unbounded (hyperbolic).  Again 
following procedures as outlined above, we get
\begin{equation}
x_n  \,=\, \cosh\sigma \,+\, \sinh\sigma \,\tanh\left[ (2n - 1) 
\frac{\sigma}{2} \right] \,-\, 1. 
\end{equation}     
\noindent In this case the solution for $y$ and $z$ would also be the same
as that given in Eq. (\ref{yz}). However, the basic properties of $x,
y$ and $z$ would change due to the unstable and unbounded dynamics,
and this is evident from the above equations. For large $n$, both $x$
and $z$ asymptotically reach a constant value that depends on the
magnitude of $\Omega$. 
\[
x_{\infty}\,=\, \lambda_{+}\, -\, 1, \;\; z_{\infty}\,
=\, x_{\infty}/\Omega^2. \]
However, $y$ would increase linearly with $n$,
when $n$ is large. 

\vspace{0.3cm}

\noindent (b) $\eta < - 2$, i.e., $\Omega^2 > 4$.

\vspace{0.3cm}

\noindent
Here both the eigenvalues of $M$ are real and negative. 
Thus we can take 
\[ \lambda_{\pm} \,=\, - \exp(\pm\sigma) \;\; \mbox{where}\;\;
\sigma \equiv \cosh^{-1}( |\eta|/2 ) \,=\, \cosh^{-1}\left( \frac{|2 - 
\Omega^2|}{2} \right).\]
 In this part the dynamics is still unstable and 
unbounded; it corresponds, 
 in the linear map, to  hyperbolic fixed point with reflections. 
On following the above
procedures, we get
\begin{equation}
x_n  \,= \,- \cosh\sigma \,-\, \sinh\sigma \,\coth\left[ (2n - 1)
\frac{\sigma}{2} \right] \,-\, 1.
\end{equation}    
\noindent The solutions for $y$ and $z$ still remain the same, 
and their behaviour for large values of $n$ are qualitatively same
as that of the previous part.

\vspace{0.3cm}

\noindent {\bf Case 3: $\eta^2 = 4$, i.e., $\Omega^2 = 0$ or $\Omega^2 = 4$}.

\vspace{0.3cm}

\noindent  These cases correspond to the marginal 
ones separating the stable and unstable motions. 
We can divide this case also into two parts. They are :

\vspace{0.3cm}

\noindent (a) $\eta = 2$, i.e., $\Omega^2 = 0.$

\vspace{0.3cm}

\noindent Here the eigenvalues of $M$ are equal, $\lambda_{\pm} = 1$ . 
The kick is not operating on the system. This implies that in the
expression of the time-evolution operator $F$, $x = y = 0$. Therefore
$F$ contains only one exponential and hence the coefficient $z$ would
increase linearly with time. This can also be seen as a limiting case;
as from Eq. (\ref{xyza}), $(x_{n+1}-x_{n}) \rightarrow -\Omega^2$ and
the Eq. (\ref{xyzc}) gives $z_n\,=\, -n$.

\vspace{0.3cm}

\noindent (b) $\eta = - 2$, i.e., $\Omega^2 = 4.$   

\vspace{0.3cm}

\noindent Again, the eigenvalues of $M$ are equal, but now $\lambda_{\pm} 
= - 1$. We can get the solution for the coefficients quite easily and these
are given by,
\begin{mathletters} 
\begin{eqnarray}
x_n &=& - \frac{ 2n + 1 }{ 2n - 1 } \,-\, 1,\\
y_n &=& - 2 \,\log | 2 n - 1 |, \\
z_n &=& \frac{x_n}{\Omega^2}.
\end{eqnarray} 
\end{mathletters}
For large $n$, the coefficients $x$ and $z$ asymptotically reach
constant values ($-2$ and $-1/2$), while the magnitude of $y$
increases logarithmically. Thus this
marginal case straddling the stable and the reflective hyperbolic
cases would have {\it power law } behaviors in time for phase space
variables.

\section {Quantum dynamics of the kicked singular oscillator}

Quantum mechanical studies of the time-dependent singular oscillator
have been carried out for some time now, for instance in \cite{camiz}
where complete analytical
solutions were given. We exploit our algebraic method to 
the special case of the kicked oscillator to lay bare properties
such as exact periodicity and quasi-periodicity. In fact, except for a 
change of terminology, the mathematics is already complete in the 
previous section. 
 
During our study of the classical nonlinear map, we had introduced the
classical time-evolution operator to show the one-to-one
correspondence between the nonlinear and the linear map.  While
classically this is not the usual approach to dynamics, in the case of
quantum dynamics, the most natural and popular way is to study the
quantum time-evolution operator $\hat{U} (t)$. Thus our classical approach 
generalizes most easily to the quantum. Previous work \cite{wang2} that points 
out exact quantum-classical correspondence in the case of the $SU(1,1)$
algebra for the coefficients of the invariant is easily understood in our
approach.

We define the operators as in the classical
case: 
\[ 
\hat{A} \,\equiv \,\frac{1}{2} \,\left( \hat{p}^2\,+\, \frac{1}
{\hat{q}^2} \right), 
\; \; \hat{B} \,\equiv\, \frac{1}{2}\,\hat{q}^2,
\;\;\mbox{ and} \;\;[\hat{A},\hat{B}] \,=\, - \frac{i}{2}\,\hbar\,
(\hat{p} \,\hat{q} \,+\,\hat{q} \,\hat{p}) \,\equiv \,2 \,\hbar \,\hat{C},\]  
where $(\hat{A}, \hat{B}, \hat{C})$ form the closed $SU(1,1)$ algebra which is 
given by $[\hat{A},\hat{B}] \,=\, 2 \,\hbar \,\hat{C}, \,[\hat{C},\hat{A}] \,=\, 
\hbar\,\hat{A}$ and $[\hat{C},\hat{B}] \,=\, -\hbar \,\hat{B}$.  
Among these operators 
$\hat{C}$ is anti-hermitian and the other two are hermitian.  In terms of these
operators our Hamiltonian would be,
\begin{equation}
{\hat{\cal H}}(t) \,=\, \hat{A} \,+\, \omega^2 \,T \,\hat{B}\,
\sum_{n} \delta( t - n T ),
\end{equation}
\noindent The time-evolution operator $\hat{U}(t)$ satisfies the equation,
\begin{equation}
i \,\hbar \,\frac{d\hat{U} ( t )}{dt} \,=\, {\hat{\cal H}}(t) \,\hat{U}(t),
\label{dUdt}
\end{equation} 
\noindent where $\hat{U}(0) \,=\, {\bf 1}$ and $\hat{U}(t)$ is unitary. 
The quantum Flouquet operator $\hat{U}(t=T)$, which we denote 
simply as $\hat{U}$, is given by 
\begin{equation}
\hat{U} \,=\, \exp\left( -\frac{i}{\hbar} \,\omega^2 \,T \,\hat{B} \right)
\exp\left( -\frac{i}{\hbar} \,T \,\hat{A} \right).
\end{equation}
\noindent For the complete dynamics, as usual, we have to determine 
the powers $\hat{U}^{n}$. Again we apply Rau's \cite{rau}
method for the derivation of the time-evolution operator at any arbitrary time.
We start with the time-evolution operator of the form,
\begin{equation}
\hat{U}(t) = \exp\left[\frac{i}{\hbar}X(t)\,\hat{B}\right]~~\exp\left[
\frac{1}{\hbar}\,Y(t)\,\hat{C}\right]~~\exp\left[\frac{i}{\hbar}\,Z(t)\,
\hat{A}\right],
\label{Ut}
\end{equation}
\noindent where $X(t), Y(t)$ and $Z(t)$ are real functions of time, so 
that $\hat{U}(t)$ remains unitary. From the
initial condition $\hat{U}(0) = {\bf 1}$, we have $X(0) = Y(0) = Z(0)
= 0$.  Substituting the above $\hat{U}(t)$ in Eq. (\ref{dUdt}),
following identical procedures as given for the classical dynamics, we
get
\begin{eqnarray}
\dot{X} &=& - X^2 - \omega^2 \,T \,\sum_{n} \delta (t - nT ), \nonumber \\
\dot{Y} &=& 2 X, \\
\dot{Z} &=& - e^{-Y}. \nonumber
\end{eqnarray}  
\noindent These set of equations for the quantum 
coefficients are {\it identical} to the equations for the classical
coefficients as given in Eq. (\ref{clco}). Therefore we would get an
identical map for the quantum coefficients as for the classical
coefficients Eq. (\ref{xyz}).  Since these two maps are identical,
their stability properties are identical, i.e., quantum dynamics
exactly follows its classical counterpart. Thus when the classical
Perron-Frobenius operator becomes identity as in Eq. (\ref{ident}) the
Flouquet operator also becomes identity.  The various cases discussed
classically, including the marginal and the reflective hyperbolic
cases have exact quantum counterparts. Thus as far as time evolution
is concerned the quantal problem is already solved. Issues of 
eigenstates and spectrum can be tackled similarly, as for example done
in \cite{BBVT} for the case of the kicked harmonic oscillator;
however we do not pursue this further here.

\section{Summary}

We have studied, both classically and quantum mechanically, a
time-dependent one-dimensional nonlinear integrable system, namely the
kicked harmonic oscillator with a singular potential. By applying our
Lie-algebraic interpretation of the LR-invariant, we have
determined the exact invariant of that nonlinear system.  One can
apply this method to determine the invariant of any time-dependent
Hamiltonian, formed by the generators of any closed algebra. We have
also found, as far as we know, a new integrable nonlinear mapping.
Here we derived that map from the Hamiltonian formed of the generators
of the $SU(1,1)$ algebra. We expect that we can determine integrable
two-dimensional mappings from the Hamiltonians formed by the
generators of other closed algebras.

We constructed the classical time-evolution operator, or the
Perron-Frobenius operator for the nonlinear integrable system, taking
advantage of a recent method of solving the time-dependent
Schr\"{o}dinger equation.  This brings into relief the extreme
quantum-classical correspondence in these systems which have algebraic
structures underlying them.  

We may speculate whether for time-dependent systems with more than one
degree of freedom, which are constructed from elements of some Lie
algebra, the nonlinear equations (whether difference or differential)
that are the analogue of the coefficient mapping in Eq. (\ref{xyz})
are integrable. Further, if there is a connection between the
integrability of these equations and that of the original system
itself. Our study that began as an attempt to study a nonlinear
time-dependent integrable system led us to one that is in many respects
related to the harmonic oscillator with time-dependent frequency. We
may then ask if we can go beyond this limitation to a more ``genuine''
form of nonlinearity.

\begin{figure}
\caption{Stroboscopic picture of the time-dependent harmonic oscillator with
a singular perturbation; the integrable behaviour is evident. All
quantities shown are dimensionless}
\label{strobo}
\end{figure}

\begin{figure}
\caption{(a) Period-9 orbit of the nonlinear map for the case
  $\Omega^2 = 2 \left[ 1 - \cos\left( \frac{2 \pi}{9} \right) \right]
  \thickapprox 0.4679$, (b) quasi-periodic orbit  for the case 
$\Omega^2 = 2.43$. All
quantities shown are dimensionless}
\label{pq}
\end{figure} 

\begin{figure}
\caption{The coefficients of the time-evolution operator showing the 
period-9 behaviour for the same $\Omega^2$ as in Fig. \ref{pq}(a).
Shown are the $x$ (a) and $y$ (b) coefficients. All
quantities shown are dimensionless}
\label{xyz1}
\end{figure}

\begin{figure}
\caption{The coefficients of the time-evolution operator showing the 
quasi-periodic  behaviour for the same $\Omega^2$ as in Fig. \ref{pq}(b).
Shown are the $x$ (a) and $y$ (b) coefficients. All
quantities shown are dimensionless}
\label{xyz2}
\end{figure}

\end{sloppypar}

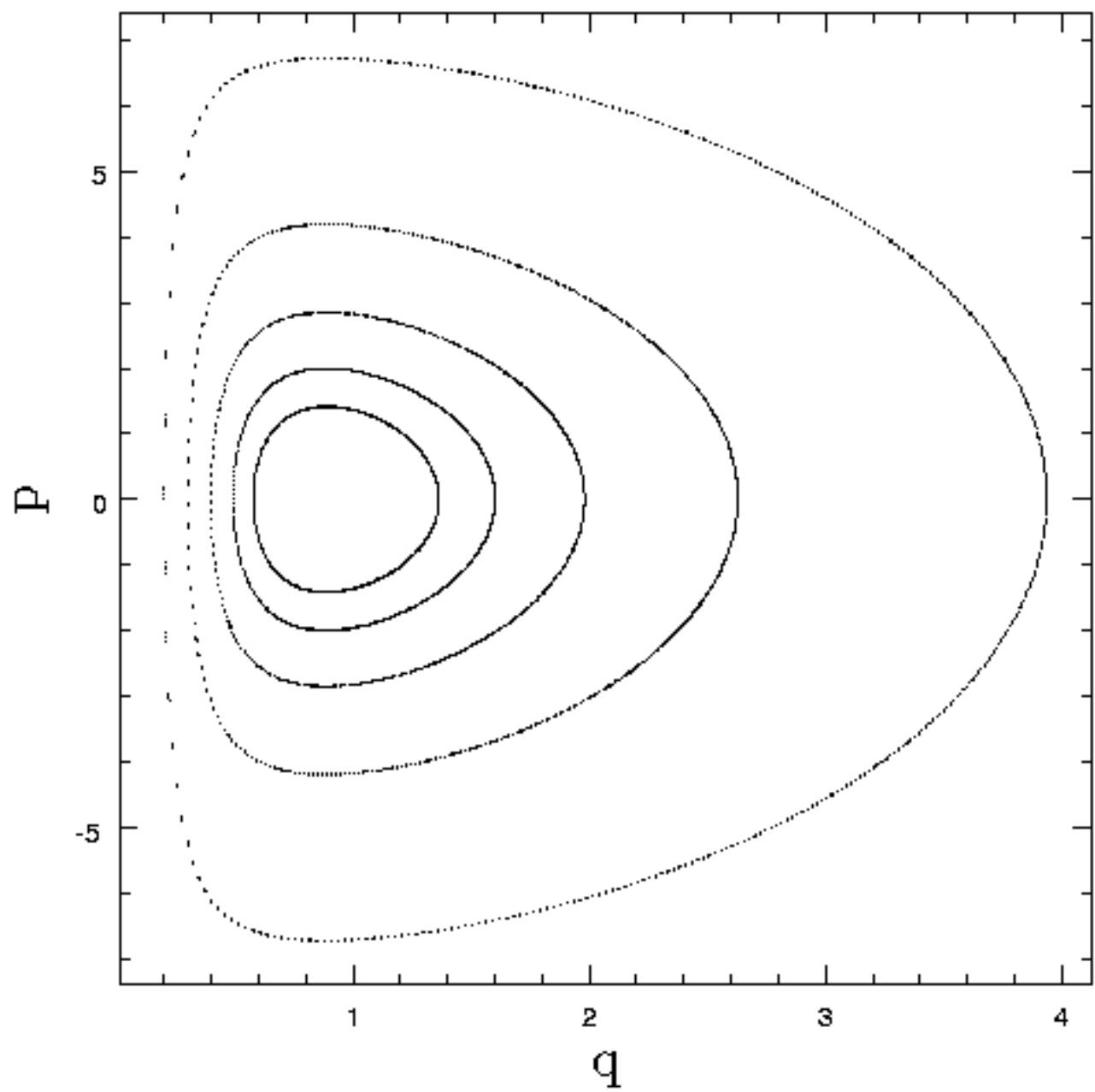
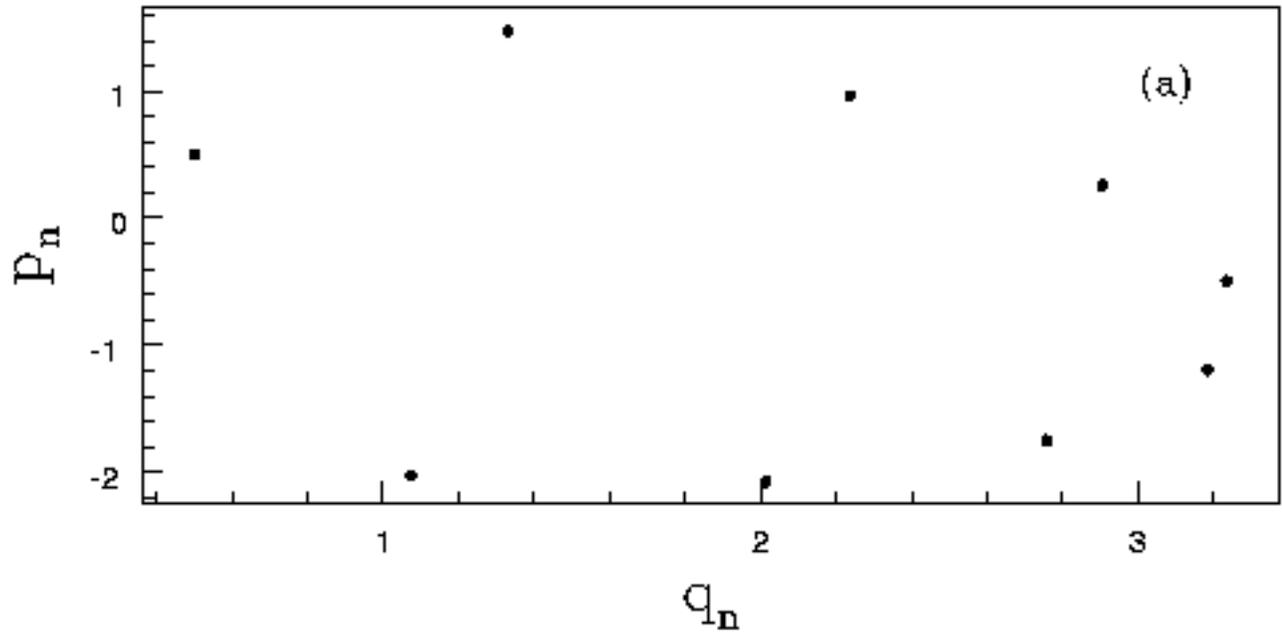
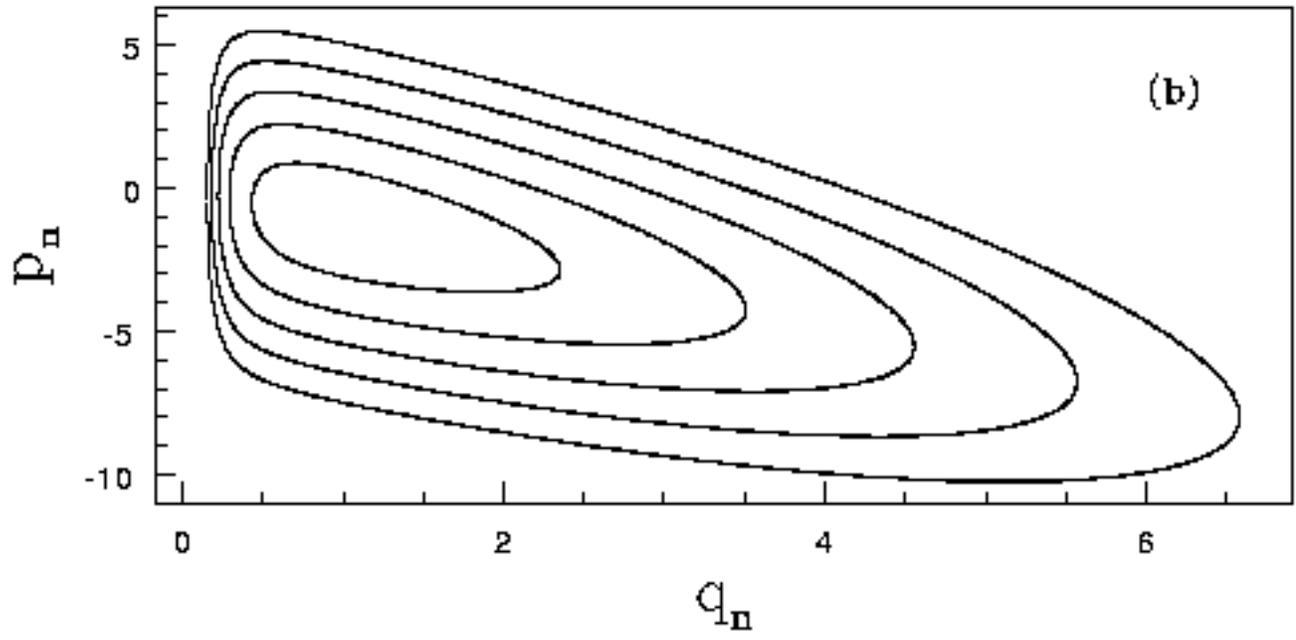
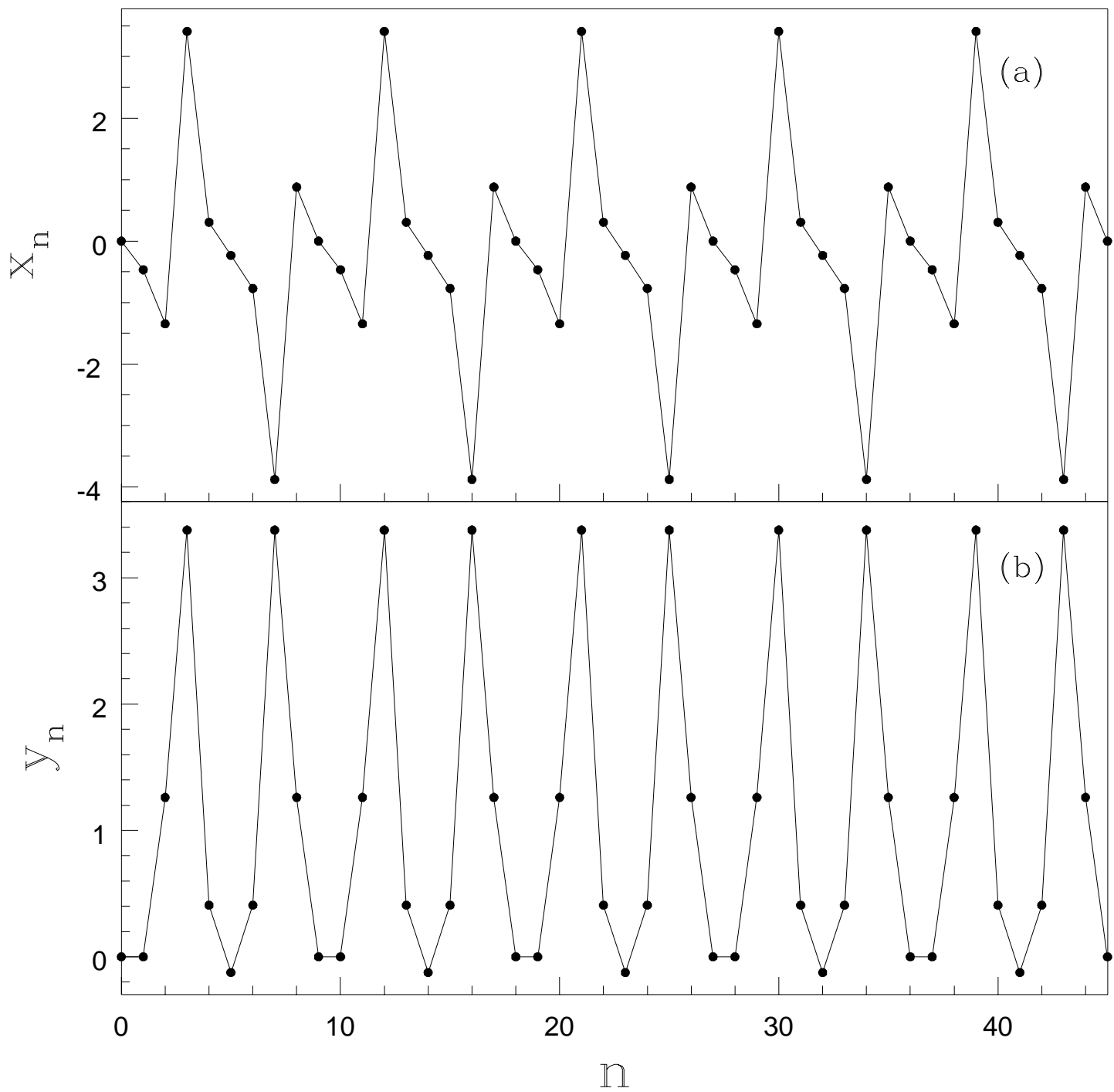
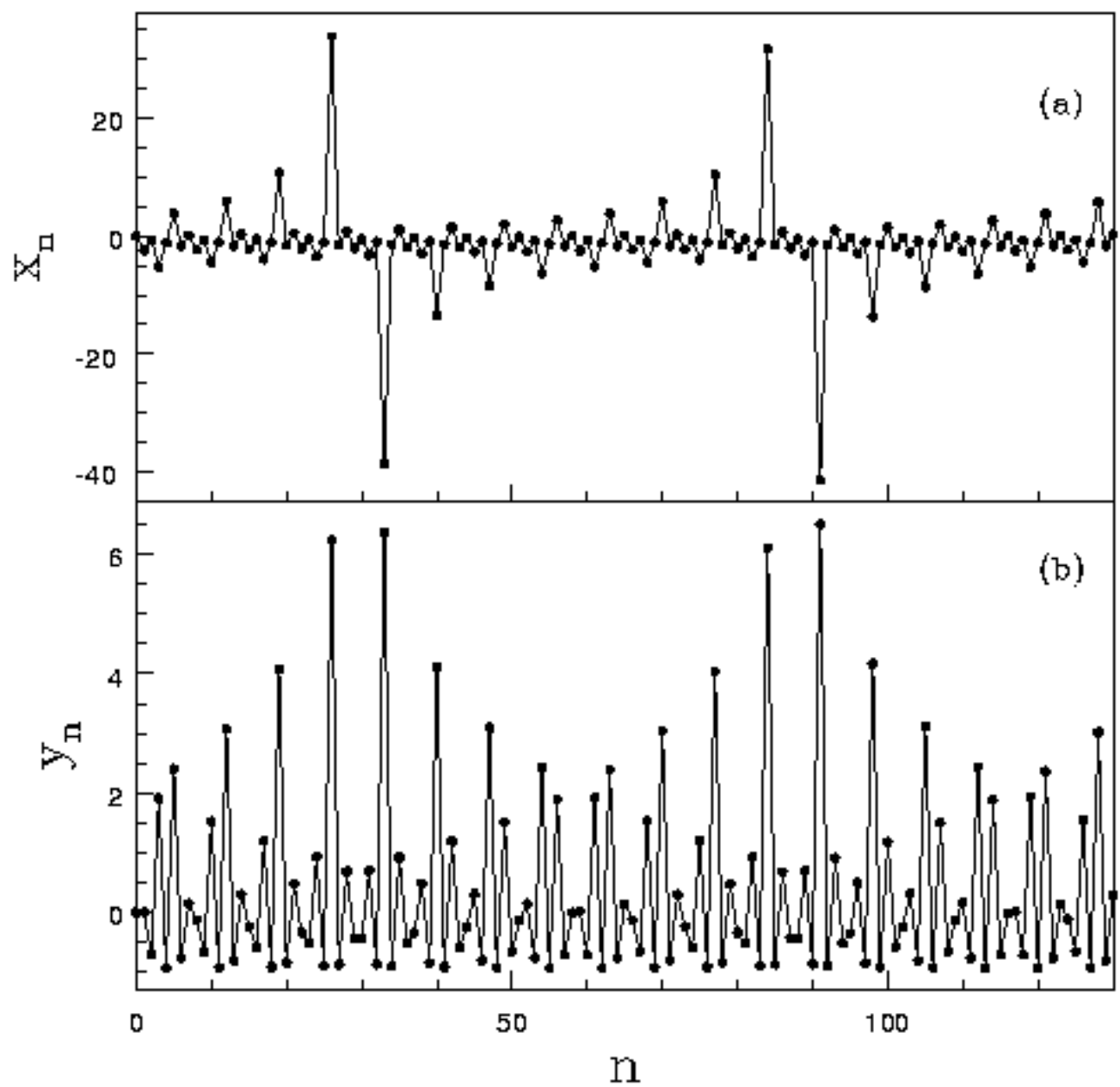
\end{document}